# Giant Negative Thermal Expansion Induced by the Synergistic Effects of Ferroelectrostriction and Spin-Crossover in PbTiO$_3$-Based Perovskites

Zhao Pan,[1,2] Jun Chen,[3,4,*] Runze Yu,[2] Lokanath Patra,[5,6] Ponniah Ravindran,[5,6] Andrea Sanson,[7] Ruggero Milazzo,[7] Alberto Carnera,[7] Lei Hu,[4] Hajime Yamamoto,[2] Yang Ren,[8] Qingzhen Huang,[9] Yuki Sakai,[2] Takumi Nishikubo,[2] Takahiro Ogata,[2] Xi'an Fan,[1] Yawei Li,[1] Guangqiang Li,[1] Hajime Hojo,[2] Masaki Azuma,[2] and Xianran Xing[3,4]

[1]State Key Laboratory of Refractories and Metallurgy, Wuhan University of Science and Technology, Wuhan 430081, China

[2]Laboratory for Materials and Structures, Tokyo Institute of Technology, 4259 Nagatsuta, Midori, Yokohama, 226-8503, Japan

[3]Beijing Advanced Innovation Center for Materials Genome Engineering, University of Science and Technology Beijing, Beijing 100083, China

[4]Department of Physical Chemistry, University of Science and Technology Beijing, Beijing 100083, China

[5]Department of Physics, Central University of Tamil Nadu, Thiruvarur 610101, India

[6]Simulation Center for Atomic and Nanoscale MATerials (SCANMAT), Central University of Tamil Nadu, Thiruvarur, Tamil Nadu, 610101, India

[7]Department of Physics and Astronomy, University of Padova, Padova I-35131, Italy

[8]X-ray Science Division, Argonne National Laboratory, Argonne, Illinois 60439, United States

[9]Center for Neutron Research, National Institute of Standards and Technology (NIST), Gaithersburg, Maryland 20899-6102, United States

*Corresponding author: junchen@ustb.edu.cn

**ABSTRACT**

The discovery of unusual negative thermal expansion (NTE) provides the opportunity to control the common but much desired property of thermal expansion, which is valuable not only in scientific interests but also in practical applications. However, most of the available NTE materials are limited to a narrow temperature range, and the NTE effect is generally weakened by means of various modifications. Here, we report an enhanced NTE effect that occurs over a wide temperature range ($\bar{\alpha}_V$ = -5.24 × 10$^{-5}$ °C$^{-1}$, 25-575 °C), and this NTE effect is accompanied by an abnormal enhanced tetragonality, a large spontaneous polarization, and a G-type antiferromagnetic ordering in the present perovskite-type ferroelectric of (1-$x$)PbTiO$_3$-$x$BiCoO$_3$. Specifically, for the composition of 0.5PbTiO$_3$-0.5BiCoO$_3$, an extensive volumetric contraction of ~4.8 % has been observed near the Curie temperature of 700 °C, which represents the highest level in PbTiO$_3$-based ferroelectrics. According to our experimental and theoretical results, the giant NTE originates from a synergistic effect of the ferroelectrostriction and spin-crossover of cobalt on the crystal lattice. The actual NTE mechanism is contrasted with previous functional NTE materials, in which the NTE is simply coupled with one ordering such as electronic, magnetic, or ferroelectric ordering. The present study sheds light on the understanding of NTE mechanisms and it attests that NTE could be simultaneously coupled with different orderings, which will pave a new way toward the design of large NTE materials.

## INTRODUCTION

Negative thermal expansion (NTE) is an interesting physical property related to the interplay among the lattice, phonons, and electrons. Materials with NTE can be used to counteract the much more common positive thermal expansion (PTE) of ordinary materials, which usually suffer from mechanical degradation and structural instability under thermal shock.[1-6] NTE materials are rare and only a very limited number of them have been discovered in recent decades.[7-13] The understanding of NTE mechanisms is still the main challenge to explore new NTE materials. Generally, thermal expansion accounts for the effects of the anharmonic lattice potential on the equilibrium lattice separations. The phonon-related NTE mechanism has been well studied in NTE framework structures such as $ZrW_2O_8$, $ScF_3$, and $Ag_3[Co(CN)_6]$.[7,8,14] However, a certain number of NTE materials exist in which the NTE is coupled with physical properties, such as magnetovolume effect in antiperovskites,[9] metal-insulator transition in $Ca_2RuO_4$,[12] valence state transformation in $LaCu_3Fe_4O_{12}$ and $BiNiO_3$,[15,16] and ferroelectrostriction in ferroelectrics.[5] Recently, studies within the NTE field have focused on exploring materials with extreme NTE characteristics. Indeed, the stronger the NTE is, the easier it is to compensate for the PTE within composite materials.[3,5]

$PbTiO_3$ (PT), is a typical perovskite-type ($ABO_3$) ferroelectric, and it is considered to be a potential mother compound to achieve a large NTE. In the last decades, PT-based ceramics such as $Pb(Zr,Ti)O_3$ and $PT-BiScO_3$ were well known for their high piezoelectric performances at the morphotropic phase boundary.[17,18] In addition, PT also exhibits a unique NTE in the perovskite family compared with other perovskites such as $BaTiO_3$, $SrTiO_3$, and $BiFeO_3$.[5] The unit cell volume of PT contracts over a wide temperature range, which extends from room temperature up to its Curie temperature ($T_C$ = 490 °C) in the ferroelectric phase, and its average intrinsic bulk coefficient of thermal expansion (CTE) is $-1.99 \times 10^{-5}$ °$C^{-1}$.[13] NTE in PT-based compounds can be controlled in the range of $-0.11 \sim -3.92 \times 10^{-5}$ °$C^{-1}$, which covers the range found in almost all other known NTE oxides.[19] The ferroelectricity plays a crucial role in the abnormal thermal expansion behavior of PT-based ferroelectrics, which has been well interpreted by the effect of spontaneous volume ferroelectrostriction (SVFS).[5] It is worth noting that the NTE in PT-based ferroelectrics is mainly attributed to the shrinkage of the polar $c$-axis. In the ferroelectric phase, the increased volume can be well maintained by the large $c/a$, which results from the strong spontaneous polarization ($P_S$). Accordingly, it is proposed that a large NTE could be obtained by improving the tetragonality ($c/a$) of PT. The flexible structure of PT allows it to achieve a large NTE by means of modulating its tetragonality with accommodating ferroelectric-active cations in the *A* and/or *B* sites.

$BiCoO_3$ and $PbTiO_3$ are isostructural, but the polar structural distortion is more pronounced in $BiCoO_3$, whose tetragonal distortion ($c/a$) of 1.27 is larger than that of PT ($c/a$ = 1.065).[20] However, $BiCoO_3$ exhibits less thermal stability than $PbTiO_3$, and the metastable nature of $BiCoO_3$ makes it decompose easily near temperatures of 460 °C.[21] It is therefore considered that the solid solutions between PT and $BiCoO_3$ can not only enhance the tetragonality of PT but also improve the thermal stability of $BiCoO_3$. Herein, we successfully achieved enhanced tetragonality and large NTE over a wide temperature range in the present binary system of $(1-x)PT-xBiCoO_3$. A combination of intriguing physical properties such as an enhanced tetragonality, a large spontaneous polarization ($P_S$), a G-type antiferromagnetism, and a giant NTE have been observed by the introduction of polar $BiCoO_3$. In particular, a colossal volumetric contraction of ~4.8% was also observed for $x = 0.5$ during the ferroelectric-to-paraelectric (FE-to-PE) phase transition, which represents the highest level ever reported in PT-based ferroelectrics. Note that the giant NTE discussed in this work is induced by an unusual synergistic effect of ferroelectrostriction and spin-crossover of cobalt on the crystal lattice, which has been well interpreted by experimental and theoretical studies.

**EXPERIMENTAL SECTION**

The samples of $(1-x)PT-xBiCoO_3$ (abbreviated as $(1-x)PT-xBC$) were prepared with a cubic anvil-type high-pressure apparatus. The stoichiometric powder mixture of PbO, $TiO_2$, $Bi_2O_3$, and $Co_3O_4$ was sealed in a gold capsule and reacted at 6 GPa and 1100 °C for 30 min. Then, 10 mg of the oxidizing agent $KClO_4$ (approximately 10 wt% of the sample) was separately added to the top and bottom of the capsule. The obtained sample was crushed and washed with distilled water to remove the remaining KCl.

The X-ray diffraction patterns of the samples were collected with a Bruker D8 ADVANCE diffractometer for phase identification. The detailed room-temperature crystal structures of all the investigated samples were extracted from the synchrotron X-ray diffraction (SXRD) data collected at the beam lines of BL02B2 ($\lambda$ = 0.419552 Å) and BL44B2 ($\lambda$ = 0.500000 Å) in SPring-8. The temperature-dependence of the crystal structure of $(1-x)PT-xBC$ was determined by the SXRD data, which were collected at beam line 11-ID-C of the Advanced Photon Source. The wavelength of synchrotron light was λ = 0.11165 Å. The detailed crystal structure was refined based on the full-profile Rietveld method by using the software FULLPROF. The initial structural model corresponds to PT (space group *P*4*mm*, NO. 99). For the characterization of the local structure distortion of (1-*x*)PT-*x*BC as a function of temperature, Co K-edge X-ray absorption near-edge fine structure (XANES) spectra were collected in transmission mode on the 1W1B beam line at the Beijing Synchrotron Radiation Facility (BSRF). The pre-edge peak intensity was determined after the pre-edge subtraction and normalization of the XAFS spectra. The temperature-dependence of the magnetic susceptibility of 0.5PT-0.5BC was measured with a SQUID magnetometer (Quantum Design, MPMS XL) in an external magnetic field of 1000 Oe. Temperature dependent neutron powder diffraction (NPD) data were collected at the NIST Center for Neutron Research on the BT-1 high resolution neutron powder diffractometer with a wavelength of 1.53980 Å. The detailed crystal and magnetic structures were refined based on the Rietveld method by using the GSAS software package.

The first-principles density functional theory (DFT) calculations were performed using the VASP code based on projected augmented plane-wave pseudopotentials.[22] The Perdew-Burke-Ernzerhof generalized gradient approximation has been considered for the treatment of exchange-correlation.[23] We chose a very large basis set of 800 eV for the plane-wave cut-off to correctly reproduce the structural parameters.[24] A fine **k** mesh grid of 8×8×8 was used, and the structural optimization was continued until the forces on the atoms converged to less than 1 meV Å$^{-1}$. To further understand the metamagnetism presented in the system, we performed fixed spin-moment calculations in which the total energy was calculated as a function of the magnetic moment.[25]

**RESULTS AND DISCUSSION**

The XRD patterns of the (1-*x*)PT-*x*BC compounds are presented in Figure 1a. The samples are of high quality with negligible impurities. All investigated samples can be well indexed into tetragonal symmetry. With the introduction of BiCoO$_3$, the (001) peak exhibits an apparent shift to the lower-angle region, which indicates the expansion of the *c*-axis. The (100) peak shows the opposite trend: it slightly shifts to a higher-angle region, which indicates the contraction of the *a*-axis. The detailed lattice parameters were refined and plotted in Figure 1b. We observed that the *c*-axis of (1-*x*)PT-*x*BC experiences an almost linearly increase, whereas the *a*(*b*)-axis shows the opposite trend, leading to an unusually enhanced tetragonality. The large tetragonality produces a pyramidal coordination rather than an octahedral coordination in the (1-$x$)PT-$x$BC solid solutions (Figure S1a and S1b). The large lattice distortion can be attributed to the large $P_S$ displacements, which were induced by the strong Pb/Bi-O hybridization and coupling interactions between the Ti/Co and Pb/Bi cations. In $AB$O$_3$ perovskite-type ferroelectrics, the $P_S$ is due to the displacements from the centroid of the oxygen polyhedrons of the $A$-site and $B$-site atoms. Here the $P_S$ displacements of the $A$-site Pb/Bi ($\delta z_A$) and $B$-site Ti/Co ($\delta z_B$) were derived from the SXRD results (Figure S2 and S3). As can be seen, both $\delta z_A$ and $\delta z_B$ show a nearly linear increase as a function of the BiCoO$_3$ content (Figure S1c). Correspondingly, the $P_S$ is gradually enhanced by the chemical substitution of BiCoO$_3$; the $P_S$ is ~59 μC cm$^{-2}$ for undoped PT,[26] ~98 μC cm$^{-2}$ for 0.5PT-0.5BC, and finally as large as ~131 μC cm$^{-2}$ for BiCoO$_3$. These $P_S$ values are much higher than typical piezoelectrics such as PbZr$_{0.52}$Ti$_{0.48}$O$_3$ (54 μC cm$^{-2}$) and PT-BiScO$_3$ (40 μC cm$^{-2}$).[17,27] The large $P_S$ provides an opportunity to explore high-performance ferroelectric materials via strain engineering by growing thin films, such as that of BiFeO$_3$.[28]

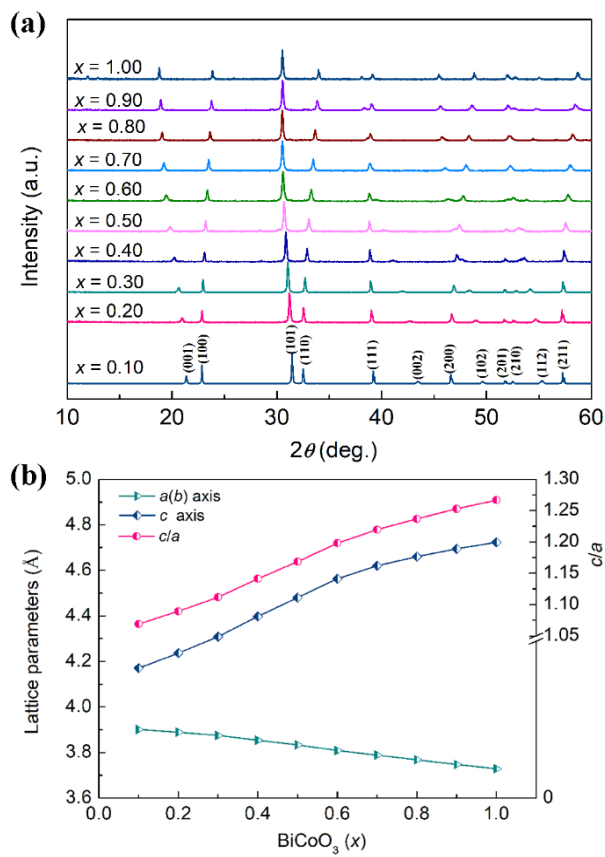

**Figure 1.** The (a) XRD patterns and (b) lattice parameters of (1-$x$)PT-$x$BC ($x$ = 0.10 ~ 1.00) at room temperature.

The large $P_S$ discussed above may be connected to a large ferroelectric volume effect, i.e. a large NTE.[5] Accordingly, the thermal expansion properties for (1-$x$)PT-$x$BC have been investigated at selected compositions (Figure S4). Detailed data for the CTE and $T_C$ of all the investigated compositions are listed in Table S1. It is very interesting to observe that both the NTE and $T_C$ are substantially enhanced by the introduction of BiCoO$_3$, up to achieve a nonlinear and strong NTE (Figure S4b). For example, let us consider the composition 0.7PT-0.3BC: in the temperature range ~25-375 °C, the unit cell volume of 0.7PT-0.3BC shows little dependence on the temperature with a small average bulk CTE $\bar{\alpha}_V$ = -6.60 × 10$^{-6}$ °C$^{-1}$. However, a very strong NTE occurs in 0.7PT-0.3BC at higher temperatures, where $\bar{\alpha}_V$ = -1.33 × 10$^{-4}$ °C$^{-1}$ in the range ~375-575 °C. In the whole temperature range ~25-575 °C, the average CTE is approximately $\bar{\alpha}_V$ = -5.24 × 10$^{-5}$ °C$^{-1}$ (Figure 2a). It needs to be mentioned that the NTE in the present 0.7PT-0.3BC is much stronger than the most representative NTE compounds, such as Invar alloys,[29] ZrW$_2$O$_8$ ($\bar{\alpha}_V$ = -2.73 × 10$^{-5}$ °C$^{-1}$),[7] ScF$_3$ ($\bar{\alpha}_V$ = -9.3 × 10$^{-6}$ °C$^{-1}$),[30] CdPt(CN)6•2{H$_2$O} ($\bar{\alpha}_V$ = -2.19 × 10$^{-5}$ °C$^{-1}$),[31] and Fe[Co(CN)$_6$] ($\bar{\alpha}_V$ = -4.41 × 10$^{-6}$ °C$^{-1}$).[32] The NTE of 0.7PT-0.3BC is also stronger than that of 0.4PT-0.6BiFeO$_3$ ($\bar{\alpha}_V$ = -3.92 × 10$^{-5}$ °C$^{-1}$) which was reported to exhibit the strongest NTE in PT-based ferroelectrics.[33] This suggests that the present (1-$x$)PT-$x$BC compounds could be effective additives to adjust the CTE of materials.

As shown in Figure S4b, with increasing content of BiCoO$_3$ NTE occurs in a narrow temperature range during the FE-to-PE phase transition. For example, an intriguingly giant volumetic contraction of ~4.8% has been observed in 0.5PT-0.5BC around the $T_C$ ~ 700 °C temperature transition (Figure 2b), concomitantly with a noticeable lattice collapse in which the $a(b)$-axis increases by 2.7% from about 3.886 to 3.991 Å, and the $c$-axis dramatically shrinks by 9.9% from about 4.429 to 3.991 Å (Figure S4a). At the same time, the $B$-site cations polyhedra change from 5-fold to 6-fold coordination due to the collapse of the $c$-axis. It is worth noting that a similar large volumetric contraction (~4.8%) under such a high temperature has never been obtained before. As a comparison, large volumetric contractions were previously obtained in Mn$_3A$N (-1.3%),[9] BiNiO$_3$ (-2.5 ~ -3.4%),[16] Pb(Ti$_{1-x}$V$_x$)O$_3$ (-1.4 ~ -3.6%),[34] and Ca$_2$RuO$_{3.73}$ (-6.7% by dilatometry).[35] For further increases in the content of $BiCoO_3$ ($x > 0.6$), the compounds decompose before the FE-to-PE phase transition temperature, which is attributed to the increased lattice distortion and weakened thermal stability of the perovskite structure (Figure S5).[36]

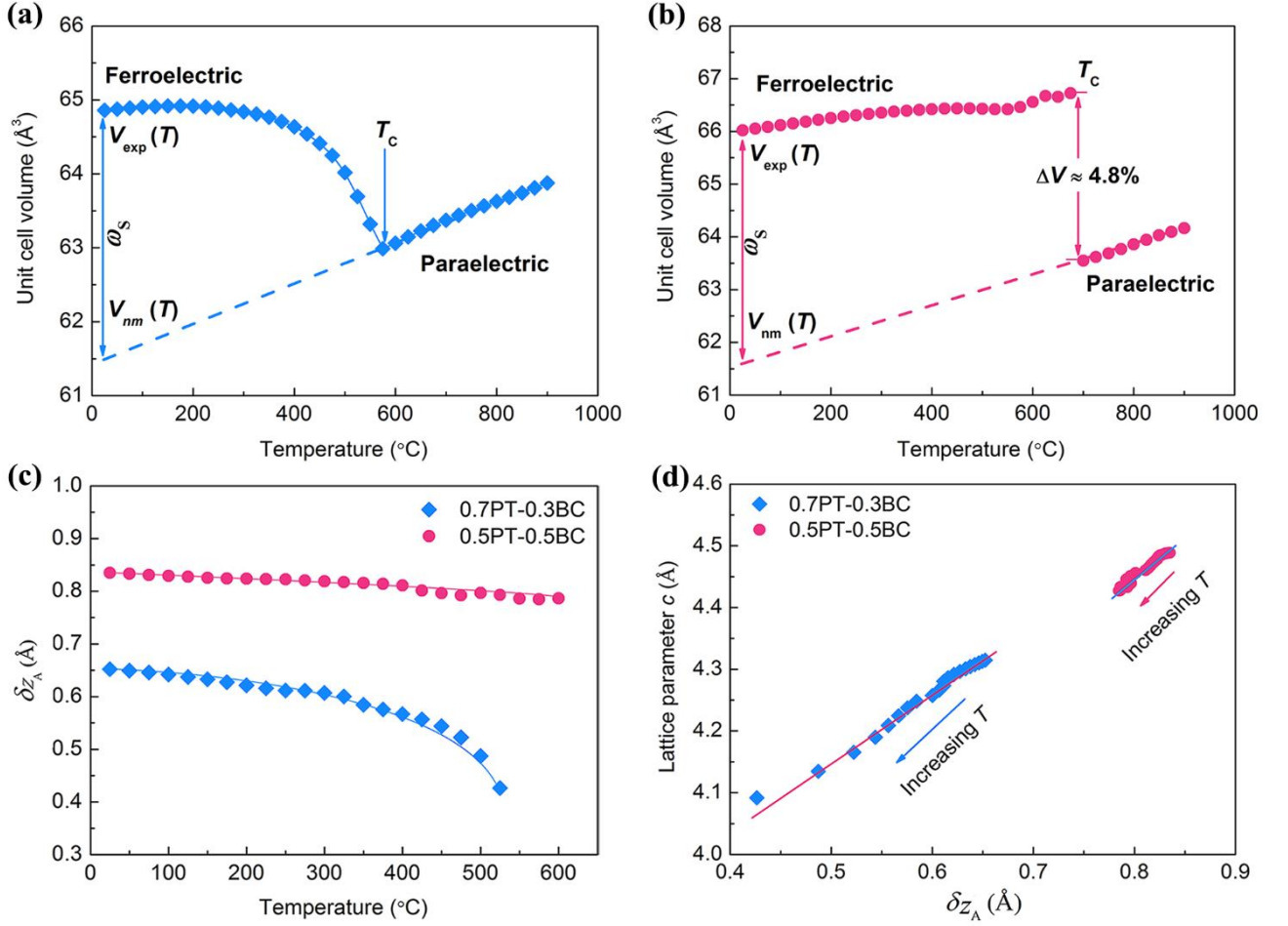

**Figure 2.** Structural data, thermal expansion, and the coupling role of (1-$x$)PT-$x$BC. Temperature-dependence of the unit cell volume for (a) 0.7PT-0.3BC and (b) 0.5PT-0.5BC, respectively. The FE-to-PE transition temperature of $T_C$ is indicated by the arrows. The unit cell volumes were obtained from the Rietveld refinements of the SXRD data. Note that the error bars are smaller than the symbols. Schematic illustration of the SVFS ($\omega_S$) is shown for the 0.7PT-0.3BC and 0.5PT-0.5BC. (c) $P_S$ displacement of $\delta z_A$ of 0.7PT-0.3BC and 0.5PT-0.5BC as a function of temperature. (d) The coupling between the $P_S$ displacement ($\delta z_A$) and the lattice parameter $c$-axis. The arrows are presented along with the elevated temperatures.

The next task involves determining the exact mechanism at the basis of the lattice contraction for (1-$x$)PT-$x$BC. Of course, the cause that determines the temperature-dependence of the $c$-axis should be the underlying reason for the NTE in (1-$x$)PT-$x$BC. Figure 2c shows the temperature-dependence of the $P_S$ displacement of the Pb/Bi atoms ($\delta z_A$) for the 0.7PT-0.3BC and 0.5PT-0.5BC compositions. A different behavior in the temperature-dependence of the $\delta z_A$ is observed. For 0.7PT-0.3BC, the $\delta z_A$ decreases continuously with increasing temperature, and it exhibits a nonlinear behavior; conversely, for 0.5PT-0.5BC, the $\delta z_A$ remains stable and exhibits a linear behavior. However, more interestingly, if we compare the overall relationship between the $c$-axis and $\delta z_A$, as reported in Figures S6 and 2d, a strong coupling interaction between the lattice ($c$-axis) and ferroelectricity ($\delta z_A$) can be found. This means that ferroelectricity determines the NTE of (1-$x$)PT-$x$BC. Specifically, when increasing temperature below the $T_C$, the polarization of 0.7PT-0.3BC decreases over a large range (Figure S7), which results in a large shrinkage of the lattice and a strong NTE. In contrast, the polarization for 0.5PT-0.5BC is well maintained in the whole ferroelectric phase, with the results of a minor change of the lattice parameters. At the FE-to-PE transition, above which ferroelectricity is lost, a giant volumetric contraction for 0.5PT-0.5BC occurs.

The NTE behavior can be well interpreted by the concept of the SVFS ($\omega_S$), which can quantitatively describe how ferroelectricity contributes to the abnormal change in the ferroelectric phase volume.[33] A large value of $\omega_S$ means a large contribution to the volume by ferroelectricity. The SVFS is defined as,

$$\omega_S = \frac{V_{exp} - V_{nm}}{V_{nm}} \times 100\%$$

where $V_{exp}$ and $V_{nm}$ are the unit cell volumes of the experimental and nominal one, respectively. The nominal one of $V_{nm}$ is basically estimated by the extrapolation from the PE to FE phase. Here, the $\omega_S$ values for 0.7PT-0.3BC and 0.5PT-0.5BC are as large as 5.4% and 7.2% (Figure 2a and 2b), respectively, which is much higher than that of PT (3.1%).[5] As shown in Figure S8, a good relationship of $\omega_S \sim \alpha \delta z_A^2$ can also be established, in which $\alpha$ is the coupling coefficient between the SVFS and polarization. Larger polarization produces a stronger effect of SVFS. From both correlations depicted in Figure 2d and Figure S8, a strong coupling between the ferroelectricity and crystal lattice is clearly achieved. Stronger ferroelectricity produces larger volumetric contraction in (1-$x$)PT-$x$BC.

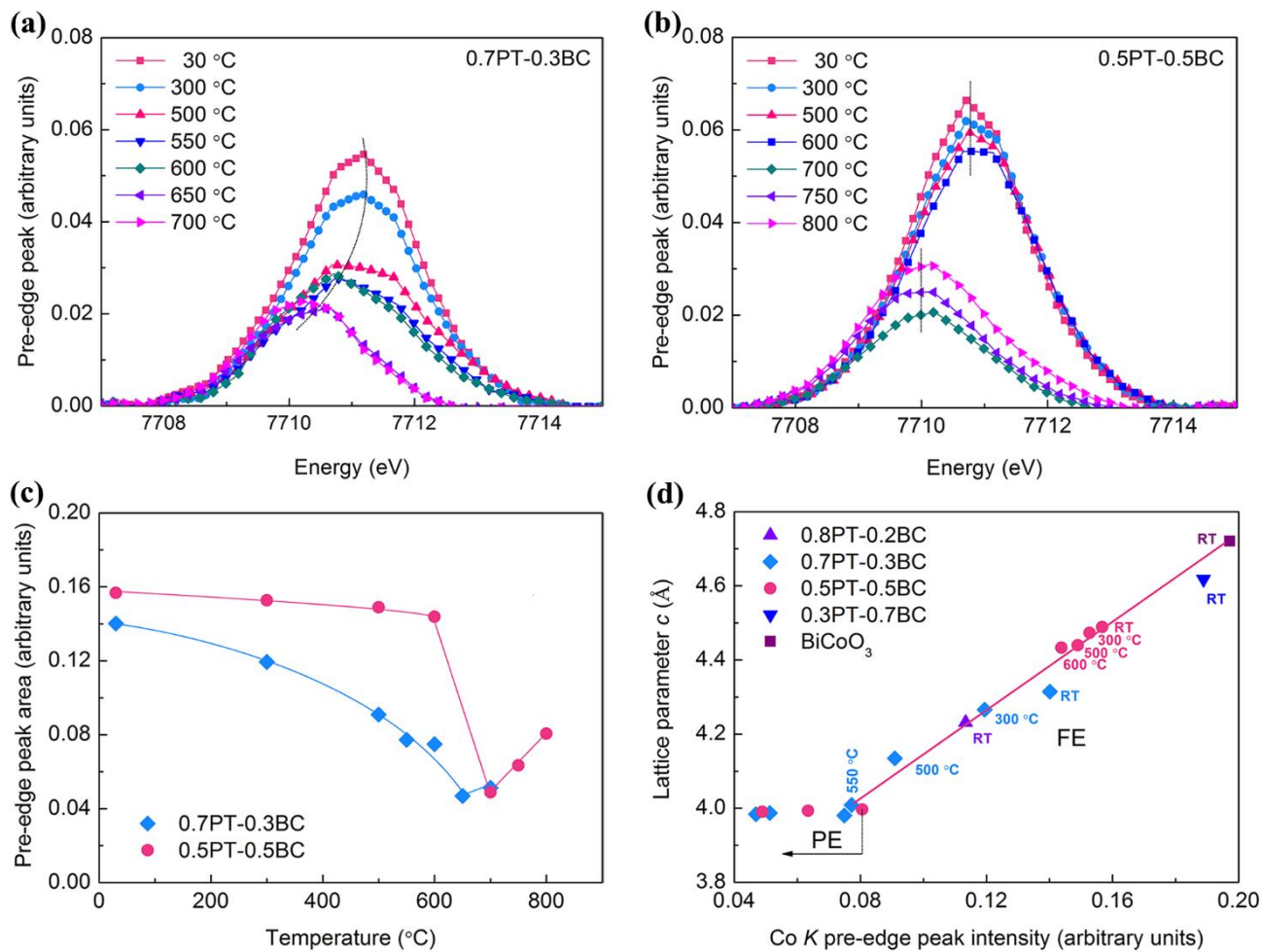

**Figure 3.** XANES characterizations of the (1-$x$)PT-$x$BC samples. Co $K$ pre-edge profiles of (a) 0.7PT-0.3BC and (b) 0.5PT-0.5BC from the normalized XANES as a function of temperature; the data were corrected after background subtraction. The dashed line indicates the energy shift. (c) Pre-edge peak intensity *vs* temperature for the 0.7PT-0.3BC and 0.5PT-0.5BC; there is a similar temperature-dependent behavior in both the pre-edge peak intensity and lattice parameters. (d) Lattice parameter $c$ *vs* Co $K$ pre-edge peak intensity for (1-$x$)PT-$x$BC as a function of composition ($x$ = 0.2, 0.3, 0.5, 0.7, and 1.0) at RT or temperature. The good linear relationship between the pre-edge peak intensity and the $c$-axis in the FE phase indicates the strong coupling between the ferroelectricity and lattice.

To further elucidate the relationship between the ferroelectricity and lattice modulation, temperature-dependent Co K-edge XANES spectra were collected in both 0.7PT-0.3BC and 0.5PT-0.5BC (Figure S9). It is well known that a pre-edge peak can be observed in the K-edge XANES of 3d transition metals such as Ti, V, Cr, and Co, which is mainly due to the electric dipole transition to the p-component in the d-p hybridization orbitals.[37] The pre-edge peak intensity increases with decreasing coordination number of the central atom and/or with increasing symmetry distortion; specifically, distortion from $O_h$ symmetry tends to dramatically enhance the pre-edge peak intensity. Therefore, a decrease in the pre-edge peak intensity in 0.7PT-0.3BC and 0.5PT-0.5BC can be ascribed to the Co displacement toward the $O_h$ symmetry or, equivalently, to the change of coordination from pyramidal $CoO_5$ to octahedral $CoO_6$, as observed in Ti and Ni species.[37] Here, a striking temperature-dependence of the pre-edge peak intensity was observed (Figure 3a and 3b). For 0.7PT-0.3BC, the pre-edge peak intensity decreases continuously with increasing temperature (Figure 3a and 3c), which indicates a gradual change of coordination from pyramidal $CoO_5$ to octahedral $CoO_6$. For 0.5PT-0.5BC, the pre-edge peak intensity remains almost constant in the ferroelectric phase; however, a sudden decrease occurs at the FE-to-PE transition (Figure 3b and 3c). Theses results indicate that the $CoO_5$ coordination of 0.5PT-0.5BC stabilizes in the ferroelectric phase and rapidly transforms to $CoO_6$ coordination during the phase transition, with Co displacement toward the $O_h$ symmetry. More interestingly, note that there is a general correlation between the Co K pre-edge peak intensity and the lattice parameter $c$ of (1-$x$)PT-$x$BC for various compositions or temperatures (Figure 3d). The pre-edge peak intensity progressively increases with increasing content of $BiCoO_3$ and decreases with elevated temperature in the FE phase (Figures S10 and 3d), this means that the degree of $CoO_5$ pyramidal coordination is progressively increased by the content of $BiCoO_3$ or by decreasing temperature. This direct correlation between the Co K pre-edge peak intensity and $c$ axis further confirms the coupling between ferroelectricity and crystal lattice.

As shown in Figure 4a, the temperature-dependence of zero-field-cooled (ZFC) and field-cooled (FC) magnetizations of 0.5PT-0.5BC suggest that it exhibits antiferromagnetic ordering below the Néel temperature ($T_N$ = ~ 60 K). The detailed magnetic structure of 0.5PT-0.5BC was determined by the NPD experiment. Long-range G-type antiferromagnetic (G-AFM) ordering was found at 5 K (Figure 4b), and no detectable antiferromagnetic peaks were observed at higher temperatures, such as 160 K (Figure 4c). The detailed refined structural parameters are summarized in Table S2. The ordered magnetic moment at 5 K is 3.24 $\mu_B$, which indicates that $Co^{3+}$ is in the HS $t_{2g}^4 e_g^2$ state in 0.5PT-0.5BC. According to previous studies, it is generally believed that $Co^{3+}$ ions adopt a low spin (LS) state in octahedral $CoO_6$ coordination environments and a high spin (HS) or an intermediate spin (IS) state in pyramidal $CoO_5$ coordination environments at low temperature and ambient pressure conditions.[21,38] The change of pyramidal $CoO_5$ coordination to octahedral $CoO_6$ could imply the transformation from the HS to the LS state of $Co^{3+}$ during the FE-to-PE phase transition in the present system of (1-x)PT-xBC. The change in unit cell volume may be accompanied by a change in the spin state (the LS $Co^{3+}$ atomic radius is smaller than that of HS $Co^{3+}$),[39] which means that the spin state of $Co^{3+}$ may be coupled with the occurrence of NTE and lattice contraction of (1-x)PT-xBC. Such a spin state transition of $Co^{3+}$ was also observed in $BiCoO_3$ during the pressure-induced FE-to-PE phase transition, in which the pyramidal coordination is stabilized by the HS electronic configuration of $Co^{3+}$ and the LS or IS state in octahedral coordination.[21]

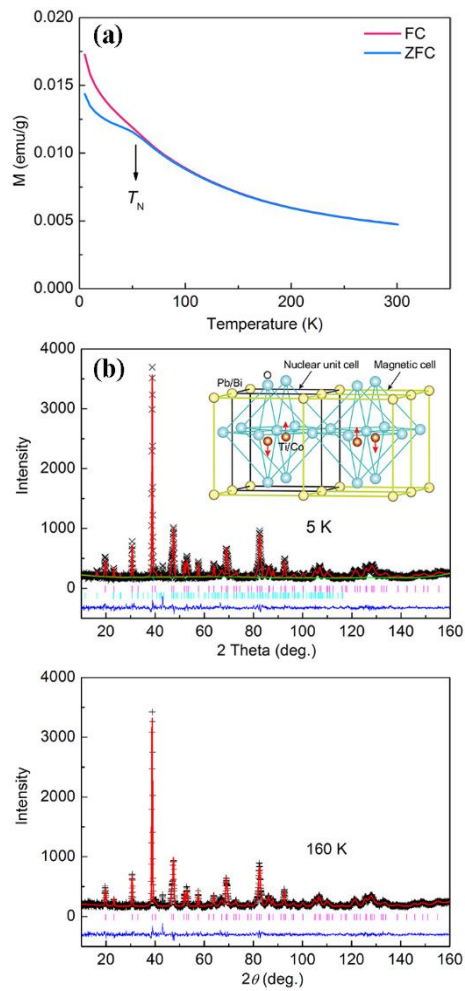

**Figure 4.** ZFC/FC magnetization and magnetic structure of 0.5PT-0.5BC. (a) ZFC/FC magnetization for 0.5PT-0.5BC in the temperature range from 5 K to 300 K in a magnetic field of 1000 Oe. Neutron powder diffraction data for 0.5PT-0.5BC at (b) 5 K and (c) 160 K. Both of the Rietveld refinements were well carried out with a tetragonal structure model. Observed (+, black), calculated (solid line, red), and difference profiles (bottom line, blue) are shown together with Bragg reflection positions. The inset shows the detailed G-type magnetic structure.
12

DFT calculations were conducted to ascertain the correlation between the crystal lattice and $Co^{3+}$ spin state. A pressure-induced metamagnetic transition was observed, wherein 0.5PT-0.5BC transforms from the HS state to the nonmagnetic LS state with > 5% volumetric collapse (Figure 5a). The obtained equilibrium volume using the Birch-Murnaghan equation of states (EOS)[40] is underestimated by only 1.6% compared with our NPD results. Moreover, our calculations show that the ground state magnetic ordering for the ferroelectric phase of 0.5PT-0.5BC is G-type, which agrees well with the NPD result. The calculated total energy for the ferroelectric phase in the nonmagnetic, ferromagnetic, A-type and C-type antiferromagnetic states are 384, 75, 63 and 7 meV (f.u.)$^{-1}$, respectively, which are higher in energy than that for the G-type state. The calculated spin moment at the Co site for the G-type state is 2.69 $\mu_B$, whereas the calculated total moment is 3.16 $\mu_B$ (f.u.)$^{-1}$; these results are in excellent agreement with the value 3.24 $\mu_B$ measured by the NPD at 5 K. A somewhat large number of magnetic moments (0.05 to 0.3 $\mu_B$) was found to reside at the O sites, which suggests that a strong d-p hybridization exists between the transition metals and O.

Likewise, the variation in the total energy as a function of the magnetic moment is shown for different unit cell volumes of 0.5PT-0.5BC in Figure 5b, as obtained from fixed spin calculations. The curves show that the energy difference between the HS and LS state is about 0.5 eV (f.u.)$^{-1}$ in the ferroelectric phase in its equilibrium volume. Approximately 5% of the volumetric compression of the total energy curve shows that the LS state is lower in energy than the HS state, which indicates that this material undergoes a pressure-induced HS-LS transition. The orbital projected density of states (DOS) for Co shows the presence of $Co^{3+}$ in the HS state in the ferroelectric phase, whereas it is in the LS state in the paraelectric phase (Figure S11). These results indicate that the HS to LS transition of $Co^{3+}$ also plays a significant important role in the giant volumetric contraction.[21,41]

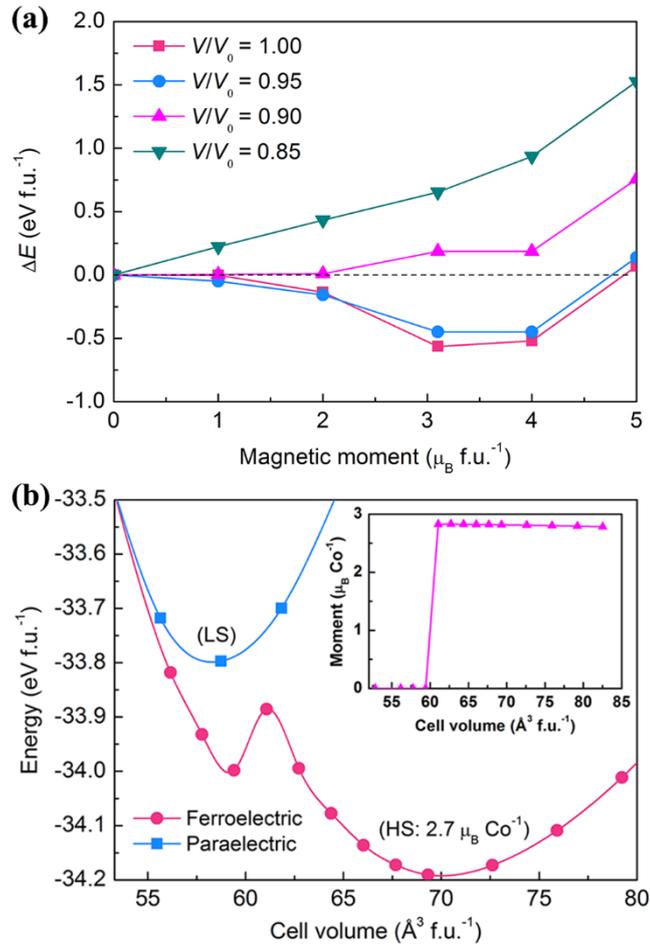

**Figure 5.** DFT calculations of 0.5PT-0.5BC. (a) Variation of total energy with magnetic moment for 0.5PT-0.5BC obtained from fixed-spin calculations for different volumes. For the equilibrium volume and 5% volumetric compression, the high-spin state with a moment of ~3.2 $\mu_B$ is found to be stable with a lower energy, while for further volumetric compression the low spin state becomes stable. (b) Calculated total energy vs. volume curves for the G-AFM ferroelectric and paraelectric phases of 0.5PT-0.5BC. The discontinuity in the total energy curve corresponds to the presence of metamagnetism in the system for which the HS equilibrium volume state changes to the LS state. The inset shows the magnetic moment at the Co site for different volumes of the G-AFM state of 0.5PT-0.5BC.

The above results demonstrate that introducing the polar perovskite of tetragonal $BiCoO_3$ into PT successfully realizes a large NTE over a wide temperature range, which is consistent with the enhanced tetragonality and ferroelectricity. Therefore, the present study may lead an effective way to achieve large NTE in PT-based ferroelectrics by introducing isostructurally strong polar perovskites to form solid solutions with PT. Other polar perovskites such as $Bi(Zn_{1/2}Ti_{1/2})O_3$ and $Bi(Zn_{1/2}V_{1/2})O_3$ also exhibit the same tetragonal structure as that of PT, while the tetragonality of these compounds are much stronger. The tetragonality of PT can be further improved in the solid solutions between PT and the aforementioned polar perovskites. Correspondingly, the related physical properties such as large NTE could be foreseeable in the strong polar PT-based perovskites, as observed in the present (1-$x$)PT-$x$BC. Another more intriguing issue is the origin of the giant volumetric collapse (~4.8 %) observed in 0.5PT-0.5BC. Note that such a large volumetric shrinkage under such a high temperature was rarely observed in previous studies. The noticeable volumetric change has been confirmed to be the synergistic effects of ferroelectrostriction and spin-crossover of cobalt on the crystal lattice. The mechanism can be in sharp contrast to previous functional NTE materials, in which NTE is always coupled with only one ordering, such as electronic, magnetic, or ferroelectric orderings.[5]

## CONCLUSIONS

In summary, a general method has been proposed to realize large NTE in PT-based ferroelectrics by the introduction of isostructural polar perovskites to further improve its tetragonality and NTE-related ferroelectricity. Enhanced NTE over a wide temperature range has been successfully achieved in the present (1-$x$)PT-$x$BC solid solutions. In particular, a giant volumetric contraction as great as of ~4.8 % has also been observed during the FE-to-PE phase transition for 0.5PT-0.5BC, which originates from the synergistic effects of ferroelectrostriction and spin-crossover of cobalt on the crystal lattice. The present work implies that the NTE could be simultaneously coupled with different orderings, which provides new insight into the understanding of NTE mechanisms and is technologically important for the exploration of compounds with large NTE.

## ASSOCIATED CONTENT

**Supporting Information**
The supporting information includes details of the refined results, the temperature-dependence of lattice constants, XANES, magnetic characterization, first-principles calculations, and other related information.

## AUTHOR INFORMATION

**Corresponding Authors**
junchen@ustb.edu.cn**Notes**

The authors declare no competing financial interests.

**ACKNOWLEDGEMENTS**

We thank Dr. L. R. Zheng (BSRF, Institute of High Energy Physics, CAS) for help with the XAFS tests and Dr. K. Kato for technical help during the high-temperature synchrotron radiation experiments at SPring-8. This work was supported by the National Natural Science Foundation of China (Grant Nos. 21805215, 21825102, and 21731001), National Program for Support of Top-notch Young Professionals, the Program for Changjiang Young Scholars, the Fundamental Research Funds for the Central Universities, China (FRF-TP-17-001B), and the General Financial Grant from the China Postdoctoral Science Foundation (2017M622536). The use of the Advanced Photon Source at the Argonne National Laboratory was supported by the U.S. Department of Energy, Office of Science, Office of Basic Energy Science (DE-AC02-06CH11357). The room-temperature synchrotron radiation experiments were performed at the BL02B2 and BL44B2 of SPring-8 with the approval of the Japan Synchrotron Radiation Research Institute (JASRI) (Proposal No. 2015B1730 and 2016A1060).

# Supplementary Information for "Giant Negative Thermal Expansion Induced by the Synergistic Effects of Ferroelectrostriction and Spin-Crossover in PbTiO$_3$-Based Perovskites"

Zhao Pan[1], Jun Chen[1,*)], Runze Yu[2], Lokanath Patra[3], Ponniah Ravindran[3], Andrea Sanson[4], Yangchun Rong[1], Lei Hu[1], Qiang Li,[1] Hajime Yamamoto[2], Linxing Zhang[1], Longlong Fan[1], Yang Ren[4], Qingzhen Huang[5], Lirong Zheng[6], Kunihisa Sugimoto[7], Yuki Sakai[9], Hajime Hojo[2], Masaki Azuma[2], Xianran Xing[1,*]

[1]*Department of Physical Chemistry, University of Science and Technology Beijing, Beijing 100083, China*

[2] *Laboratory for Materials and Structures, Tokyo Institute of Technology, 4259 Nagatsuta, Midori, Yokohama, 226-8503, Japan*

[3]*Department of Physics, Central University of Tamil Nadu, Thiruvarur, 610101, India*

[4]*Department of Physics and Astronomy, University of Padova, Padova I-35131, Italy*

[5]*X-Ray Science Division, Argonne National Laboratory, Argonne, Illinois 60439, United States*

[6]*Center for Neutron Research, National Institute of Standards and Technology (NIST), Gaithersburg, Maryland 20899-6102, United States*

[7]*Beijing Synchrotron Radiation Facility, Institute of High Energy Physics, Chinese Academy of Sciences, Beijing 100039, China*

[8]*Japan Synchrotron Radiation Research Institute, 1-1-1 Kouto, Sayo, Hyogo 679-5198, Japan*

[9]*Kanagawa Academy of Science and Technology, KSP, 3-2-1 Sakado, Takatsu-ku, Kawasaki City, Kanagawa, 213-0012, Japan*

---

*)Author to whom correspondence should be addressed. Electronic mail: junchen@ustb.edu.cn or xing@ustb.edu.cn.

**Methods**

**Sample preparation.** All samples were prepared with a cubic anvil-type high-pressure apparatus. Stoichiometric mixture powder of PbO, $TiO_2$, $Bi_2O_3$, and $Co_3O_4$ was sealed in a gold capsule and reacted at 6 GPa and 1373 K for 30 min. A 10 mg amount of the oxidizing agent $KClO_4$ (about 10 wt% of the sample) was added to the top and bottom of the capsule in a separate manner. The obtained sample was crushed and washed with distilled water to remove the remaining KCl.

**Crystal structure determination.** The room-temperature crystal structures of all investigated samples were extracted from SXRD data collected at beam line BL02B2 of SPring-8 with wavelength $\lambda = 0.419552$ Å. The temperature dependence of crystal structure of 0.7PT-0.3BC and 0.5PT-0.5BC were determined by SXRD data which were collected at beam line 11-ID-C of the Advanced Photon Source. The wavelength of synchrotron light was $\lambda = 0.11165$ Å. The detailed crystal structure was refined based on the full-profile Rietveld method performed on the software FULLPROF. The initial structural model corresponds to PT (space group *P4mm*, NO. 99).

**X-ray absorption fine structure (XAFS).** As for the characterization of the local structure distortion of (1-*x*)PT-*x*BC, XAFS spectra were collected on the 1W1B beam line at Beijing Synchrotron Radiation Facility (BSRF). All samples were recorded at the Co *K*-edge ($E = 7709$ eV) with transmission mode. The XAFS data processing and fitting were performed by using the IFEFFIT program.

**Magnetic measurement.** The temperature dependence of the magnetic susceptibility (ZFC/FC) of 0.5PT-0.5BC was measured with a SQUID magnetometer (Quantum Design, MPMS XL) in an external magnetic field of 1000 Oe. Temperature dependent NPD data was collected at the NIST Center for Neutron Research on the BT-1 high resolution neutron powder diffractometer with a wavelength of 1.53980 Å. The detailed crystal and magnetic structures were refined based on the Rietveld method by using the software GSAS.

**First-principle calculations.** Our first-principles DFT calculations were performed using the VASP[31] based on projected augmented plane wave pseudopotentials. Perdew-Burke-Ernzerhof generalized gradient approximation[32] has been considered for the treatment of exchange-correlation. We chose a very large basis set of 800 eV for the plane wave cut-off to reproduce the structural parameters correctly[33]. A fine **k**-mess grid of 8×8×8 was used and structural optimization were continued until the forces on the atoms converged to less than 1 meV Å$^{-1}$. In order to understand the metamagnetism present in the system, we have performed fixed spin moment calculations[34] in which the total energy is calculated as a function of magnetic moment.

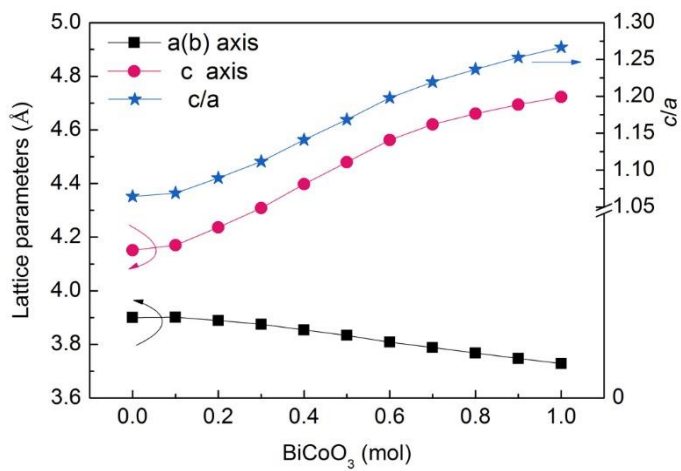

**Extended Data Figure 1 | The evolution of lattice parameters of (1-$x$)PT-$x$BC as a function of $x$.** The $c$ axis shows an almost linearly increasing tendency, while the $a$ axis shows the opposite trend. As a result, tetragonality ($c/a$) is enhanced by chemical substitution of BiCoO$_3$.

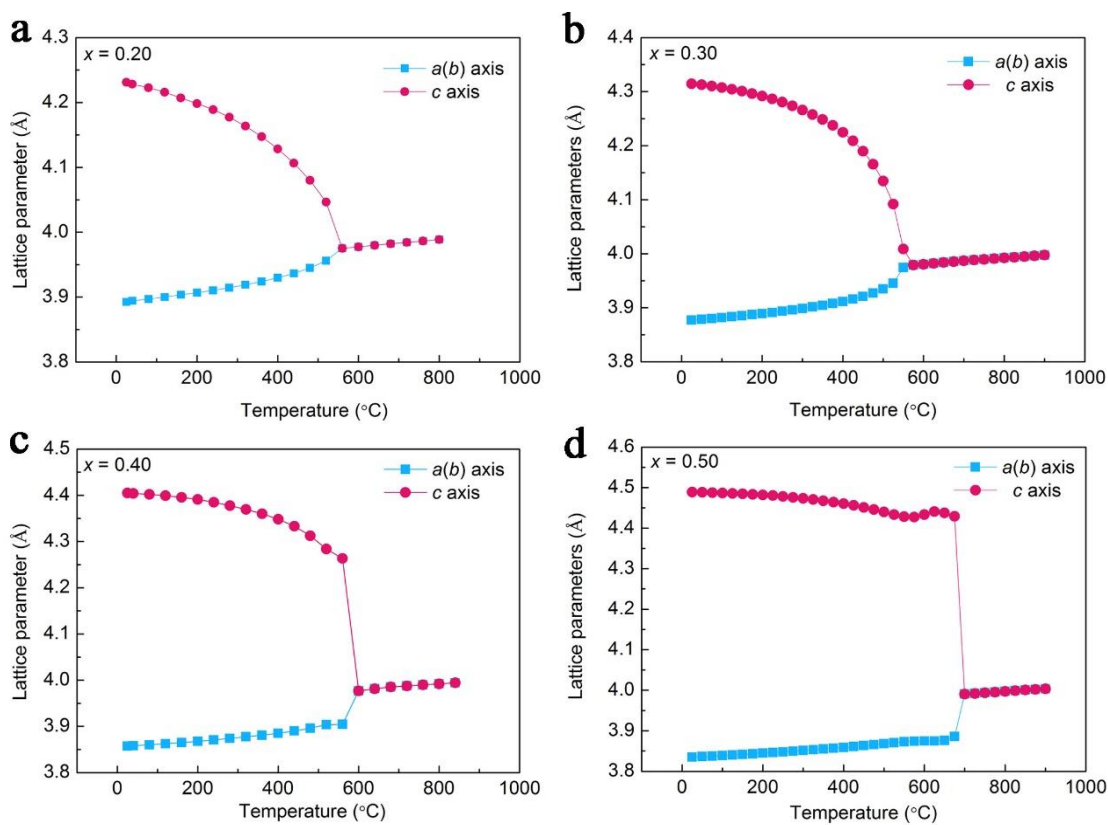

**Extended Data Figure 2 |** Temperature dependence of lattice parameters for (1-$x$)PT-$x$BC ($x$ = 0.2, 0.3, 0.4, and 0.5).

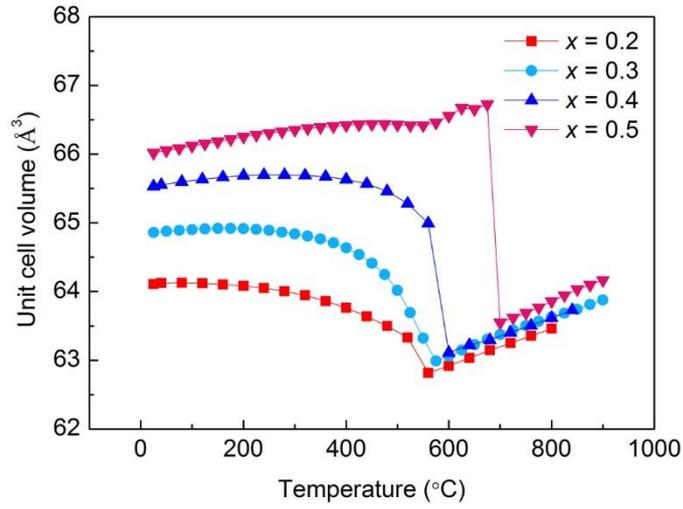

**Extended Data Figure 3** | Temperature evolution of unit cell volume for (1-$x$)PT-$x$BC ($x$ = 0.2, 0.3, 0.4, and 0.5).

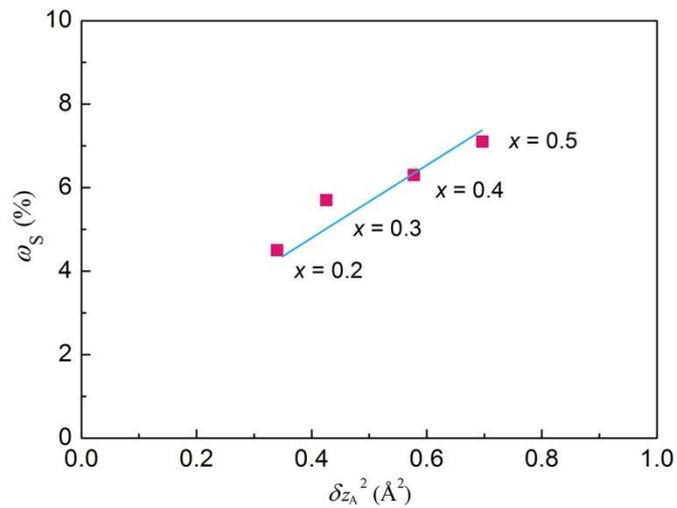

**Extended Data Figure 4** | **The correlation between the spontaneous volume ferroelectrostriction ($\omega_S$) and $P_S$ displacement ($\delta z_A$).** There is a strong correlation of $\omega_S = \alpha \delta z_A^2$ for (1-$x$)PT-$x$BC ($x$ = 0.2, 0.3, 0.4, and 0.5) ferroelectrics.

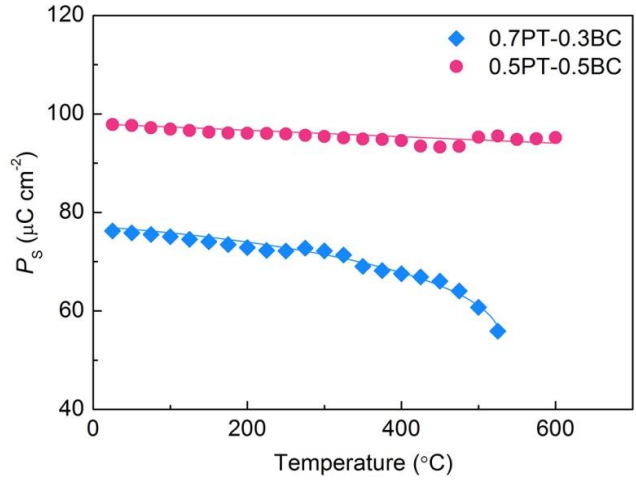

**Extended Data Figure 5 |** Temperature dependence of calculated $P_S$ for 0.7PT-0.3BC and 0.5PT-0.5BC.

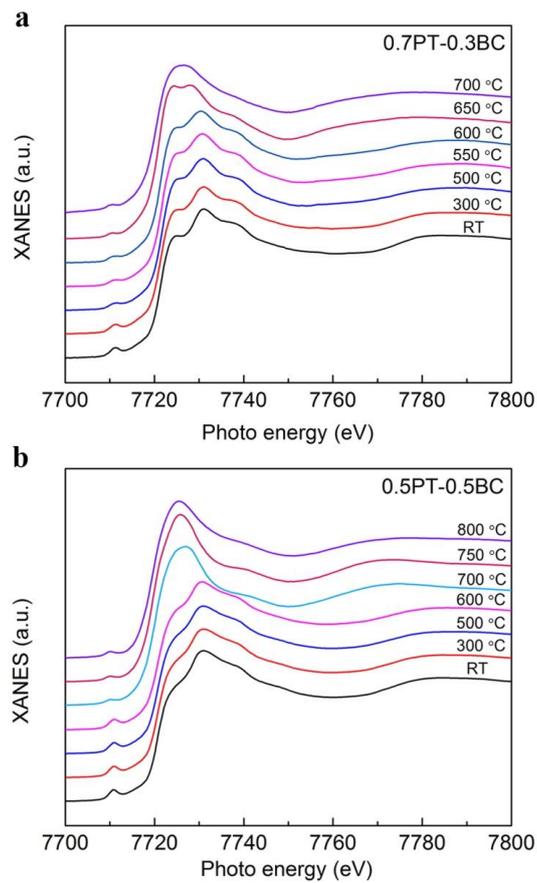

**Extended Data Figure 6 | XANES structure analysis.** Temperature dependence of XANES profiles at Co $K$-edge for **(a)** 0.7PT-0.3BC and **(b)** 0.5PT-0.5BC.

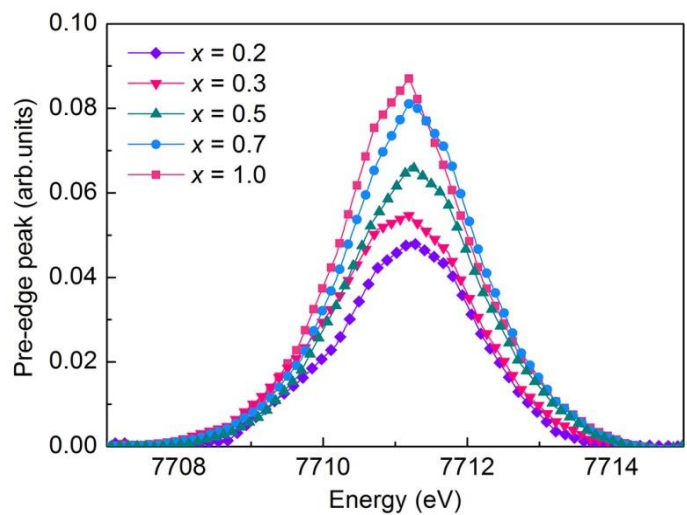

**Extended Data Figure 7 | Composition dependence of Co *K* pre-edge peak for (1-*x*)PT-*x*BC (*x* = 0.2, 0.3, 0.5, 0.7, and 1.0) at RT.**

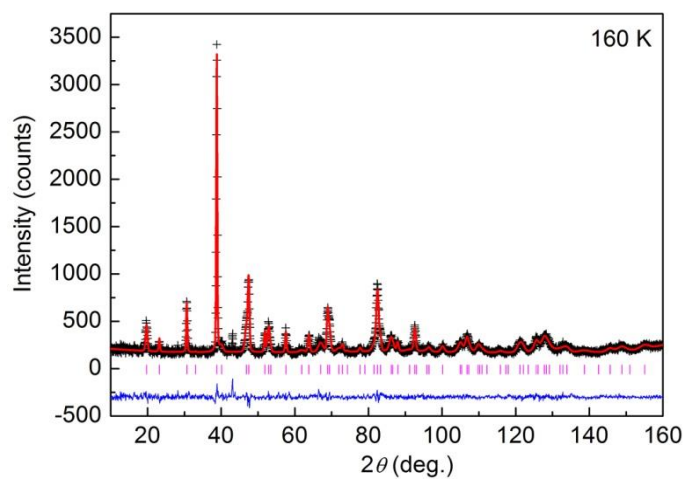

**Extended Data Figure 8 | NPD data for 0.5PT-0.5BC at 160 K.** Observed (＋, black), calculated (solid line, red), and difference profiles (bottom line, blue) are shown together with Bragg reflection positions.

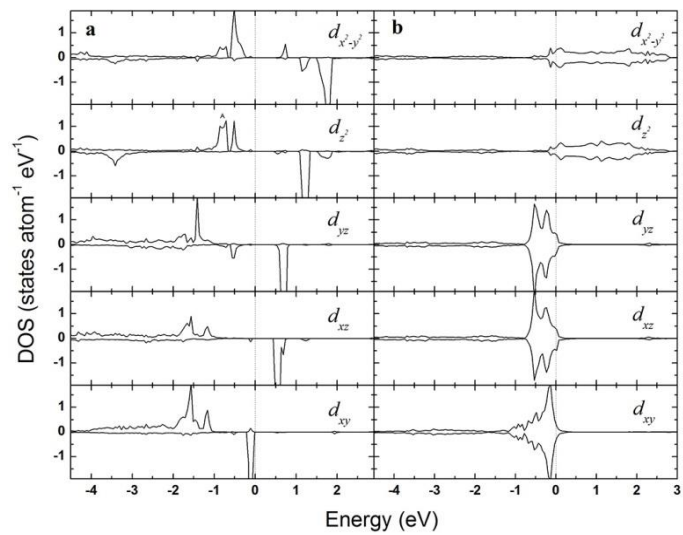

**Extended Data Figure 9** | Calculated orbital projected DOS for 0.5PT-0.5BC in the ferroelectric (left) and the paraelectric (right) phases.

**Extended Data Table 1 | The data of the CTE, volume contraction, and $T_C$ of (1-$x$)PT-$x$BC.**

| content | $x = 0$ | $x = 0.2$ | $x = 0.3$ | $x = 0.4$ | $x = 0.5$ |
|---|---|---|---|---|---|
| $T_C$ (°C) | 490 | 560 | 575 | 600 | 700 |
| CTE (×10$^{-5}$/°C) | -1.99 | -3.78 | -5.24 | —— | —— |
| $\Delta V$ | -1.29% | -2.04% | -2.97% | -3.24% | -4.76% |

**Extended Data Table 2 | Structural refinement results of NPD of 0.5PT-0.5BC at 5 K.** O1 and O2 were refined by using anisotropic atomic displacement parameters ($U_{11}$, $U_{22}$, and $U_{33}$). Others are isotropic ones. The quality of the agreement between the observed and calculated profiles is indicated by the weighted $R_{wp}$ and the goodness of fit $\chi^2$.

| atom | site | $x$ | $y$ | $z$ | $U_{iso}/U_{ij}$ |
|---|---|---|---|---|---|
| Pb/Bi | 1a | 0 | 0 | 0 | 0.02177 |
| Co/Ti | 1b | 0.5 | 0.5 | 0.597305 | 0.02456 |
| O1 | 1b | 0.5 | 0.5 | 0.178456 | 0.03044($U_{11}$)<br>0.03044($U_{22}$)<br>0.01287($U_{33}$) |
| O2 | 2c | 0.5 | 0 | 0.684938 | 0.01585($U_{11}$)<br>0.00963($U_{22}$)<br>0.02077($U_{33}$) |

*Space group $P4mm$, $a(b)$ = 3.82055(4) Å, $c$ = 4.49313(2) Å, $m_z$ (Co) = 3.24 µB, $\chi^2$ = 1.28, $R_{wp}$ = 6.82%.